\documentclass[fleqn,showpacs,preprint,preprintnumbers,amsmath,amssymb]{revtex4}

\usepackage{graphicx}
\usepackage{dcolumn}
\usepackage{bm}

\begin{document}

\title{Dynamics of nonlinear ion-waves in Fermi-Dirac electron-positron-ion magnetoplasmas}

\author{M. Akbari-Moghanjoughi}
\affiliation{Azarbaijan University of
Tarbiat Moallem, Faculty of Sciences,
Department of physics, 51745-406, Tabriz, Iran}

\date{\today}

\begin{abstract}
Oblique propagation and head-on collisions of solitary structures is studied in a dense magnetized plasma comprised of relativistic ultra-cold electrons and positrons and positive dynamic ions using conventional extended multi-scales technique, in the ground of quantum hydrodynamics model. The variations of head-on collision phase-shift as well as the characteristic soliton amplitude and width is evaluated numerically in terms of other plasma parameters such as mass-density, normalized magnetic field strength, its angle with respect to the soliton propagation and the relative positron number-density. The relevance of current investigations, with appropriate plasma parameters for the astrophysical dense magnetized objects such as white-dwarfs, is addressed.
\end{abstract}


\maketitle

\section{Background}\label{intro}

Extreme conditions in plasma such as high temperatures and pressures can lead to different features of nonlinear wave dynamics, including propagation and collision, from those encountered in ordinary ionized environments \cite{akbari1}. It is well known that dispersive dynamics arise due to quantum diffraction phenomenon in many of degenerate plasma kinds \cite{haas, shukla, manfredi}. It has also been confirmed that in a degenerate plasma there may be critical values which govern the nonlinear dynamics \cite{akbari2}. The degeneracy feature which is a fundamental aspect of ordinary solids arise due to exclusion mechanism when the de Broglie thermal wavelength $\lambda_B = h/(2\pi m_e k_B T)^{1/2}$ is comparable or higher than inter-particle distances \cite{bonitz} and leads to a much higher pressure on degenerated species compared to the non-degenerated ones. It has been shown \cite{chandra1} that in a completely degenerate astrophysical object such as a white dwarf the electrons, which follow Fermi-Dirac statistics rather than that of Boltzmann, become relativistically degenerate due to gigantic inward gravitational pressure and this can lead to softening of the degeneracy pressure giving rise to the ultimate collapse of the star. The effects of the relativistic degeneracy has recently been considered in electron-positron-ion plasmas on large-amplitude nonlinear wave dynamics by number of authors in ultra-relativistic and non-relativistic degeneracy limits \cite{mamun, akbari3}.

On the other hand, more recently, the dynamics of nonlinear ion waves has been investigated in electron-ion plasma considering a wide range of relativistic degeneracy and it has been confirmed that the relativity parameter plays a crucial role in propagation of such waves in dense Fermi-Dirac plasmas \cite{akbari4}.\textbf{ However, to our current knowledge, effect of the relativistic degeneracy parameter, $R$ (defined in next section), which is directly related to quantities such as the electron number-density, $n$, and mass-density of plasma, $\rho$, has not been considered in collision dynamics of solitary structure.} Historically, discovery of the remarkable shape-preservation feature of these waves during interactions by Zabusky et.al. \cite{zabusky} in 1965, made their first important applications in communication technology. Later, Washimi et.al \cite{washimi} showed that such waves, in a weakly nonlinear regime, can be mathematically modeled by the well known Korteweg-de Vries (KdV) equations and Oikawa et.al. \cite{oikawa} have used an extended approach to consider the interaction of such waves as the superposition of two single KdV-type solitons.

Electron-positron-ion plasmas are important because they appear in many astrophysical environment such as, active-galactic-nuclei, pulsar magnetospheres, neutron stars and supernovas etc. \cite{shapiro, rees, miller, goldr, michel}. They may even play important role in \textbf{evolution of the early Universe} \cite{silva}. Electron-positron-ion plasmas can also be produced in intense-laser matter interaction experiments \cite{surko1, surko2, greeves, Berezh}. It has been noted \cite{akbari1} that the characteristic nonlinear wave frequencies are comparably higher compared to that of pair annihilation in such dense plasmas so that these waves \textbf{survive in these extreme environments}. In the current study we consider in a magnetized electron-positron-ion plasma the propagation and head-on collisions of ion-acoustic solitary waves in a wide range of relativity parameter, i.e. plasma mass-density, thus the findings may well be applicable to inertial confinement fusion electron-positron-ion plasmas. The article is organized in the following way. The basic normalized hydrodynamics equations are introduced in section \ref{basic}. Evolution equations along with collision parameters are derived in section \ref{shift}. Numerical analysis and \textbf{discussion are} presented in section \ref{discussion} and final remarks are drawn in section \ref{conclusion}.

\section{Quantum Hydrodynamic Fluid Model}\label{basic}

Considering a quantum fully degenerate plasma with \textbf{relativistically degenerate} electrons/positrons, we will use the conventional quantum hydrodynamic (QHD) fluid equations to describe the dynamics of nonlinear excitations. The pair-annihilation rate is ignored compared to characteristic plasma frequencies \cite{akbari1} and the plasma is considered as collision-less due to Pauli-blocking mechanism. Therefore, the closed set of QHD equations may be written in the following dimensional form
\begin{equation}\label{dim}
\begin{array}{l}
\frac{{\partial {n_i}}}{{\partial t}} + \nabla  \cdot \left( {{n_i}{{\bf{V}}_i}} \right) = 0,\hspace{3mm}{{\bf{V}}_i} = {\bf{i}}{u_i} + {\bf{j}}{v_i} + {\bf{k}}{w_i}, \\ \frac{{\partial {n_j}}}{{\partial t}} + \nabla  \cdot \left( {{n_j}{{\bf{V}}_j}} \right) = 0,\hspace{3mm}{{\bf{V}}_j} = {\bf{i}}{u_j} + {\bf{j}}{v_j} + {\bf{k}}{w_j}, \\ \frac{{\partial {{\bf{V}}_i}}}{{\partial t}} + \left( {{{\bf{V}}_i} \cdot \nabla } \right){{\bf{V}}_i} =  - \frac{e}{{{m_i}}}\nabla \phi  - \frac{{\gamma {k_B}{T_i}}}{{{m_i}{n_{i0}}}}{\left( {\frac{{{n_i}}}{{{n_{i0}}}}} \right)^{\gamma  - 2}}\nabla {n_i} + B_0({{\bf{V}}_i} \times {\bf{k}}), \\ \frac{{{m_j}}}{{{m_i}}}\left( {\frac{{\partial {{\bf{V}}_j}}}{{\partial t}} + \left( {{{\bf{V}}_j} \cdot \nabla } \right){{\bf{V}}_j}} \right) =  - \frac{{{e s_j}}}{{{m_i}}}\nabla \phi  - \frac{1}{{{m_i}{n_j}}}\nabla {P_j} + \frac{{{m_j}}}{{{m_i}}}\left( {\frac{{{\hbar ^2}}}{{2m_j^2}}} \right)\nabla \left( {\frac{{{\nabla ^2}\sqrt {{n_j}} }}{{\sqrt {{n_j}} }}} \right), \\ {\nabla ^2}\phi  =  - 4\pi e\left( {\sum\limits_j {{s_j}{n_j}}  + {n_i}} \right), \\
\end{array}
\end{equation}
where, $\textbf{B}= B_0\textbf{k}$, $j=\{e,p\}$ and $s_j=\{-1,+1\}$, for electrons and positrons, respectively, and $\hbar$ is the normalized Plank constant. It is noted that, in fully-degenerate configuration from Fermi-Dirac statistical model, the relativistic degeneracy pressure is expressed in the following general form, which is valid to the electrons/positrons with arbitrary degree of relativistic degeneracy \cite{chandra2}
\begin{equation}
{P_j} = \frac{{\pi m_j ^4{c^5}}}{{3{h^3}}}\left\{ {{R_j}\left( {2{R_j^{2}} - 3} \right)\sqrt {1 + {R_j^{2}}}  + 3\ln\left[ {R_j + \sqrt {1 + {R_j^{2}}} } \right]} \right\},
\end{equation}
\textbf{in which the relativity parameter $R_j=(n_j/n_0)^{1/3}$ \cite{akbari7} (${n_0} = \frac{{8\pi m_\alpha^3{c^3}}}{{3{h^3}}}\simeq 5.9 \times 10^{29} cm^{-3}$) is the ratio of electron/positron Fermi relativistic momentum $p_{Fj}$ to $m_e c$.} Thus, with the new relativistic degeneracy pressure term the basic equations may be rewritten as
\begin{equation}\label{normal}
\begin{array}{l}
\frac{{\partial {n_i}}}{{\partial t}} + \nabla  \cdot \left( {{n_i}{{\bf{V}}_i}} \right) = 0, \\
\frac{{\partial {n_j}}}{{\partial t}} + \nabla  \cdot \left( {{n_j}{{\bf{V}}_j}} \right) = 0, \\
\frac{{\partial {{\bf{V}}_i}}}{{\partial t}} + \left( {{{\bf{V}}_i} \cdot \nabla } \right){{\bf{V}}_i} = - \frac{e}{{{m_i}}}\nabla \phi - \left( {\frac{{\gamma {k_B}{T_i}}}{{n_{i0}{m_e}{c^2}}}} \right)\left(\frac{n_i}{n_{i0}}\right)^{\gamma  - 2}\nabla {n_i} + B_0({{\bf{V}}_i} \times {\bf{k}}), \\ \frac{{{m_j}}}{{{m_i}}}\left( {\frac{{\partial {{\bf{V}}_j}}}{{\partial t}} + \left( {{{\bf{V}}_j} \cdot \nabla } \right){{\bf{V}}_j}} \right) = -\frac{{{e s_j}}}{{{m_i}}}\nabla \phi - \nabla\left[ {1 + {R_0^{2}\alpha_j^{2/3} {n_j}^{2/3}}} \right]^{1/2} + {\frac{{{m_j}}}{{{m_i}}}}\left( \frac{{{H_r^{2}}}}{2}\right)\nabla \left( {\frac{{{\nabla ^2}\sqrt {{n_j}} }}{{\sqrt {{n_j}} }}} \right), \\
{\nabla ^2}\phi = - 4\pi e\left( {\sum\limits_j {{s_j}{n_j}}  + {n_i}} \right), \\
\end{array}
\end{equation}
where, $R_0=(n_{e0}/n_0)^{1/3}$ ($\alpha_j=\{1, \alpha\}, \alpha=n_{p0}/n_{e0}$ for electrons and positrons, respectively) is a measure of the relativistic effects (called the relativity parameter) and $H_r^{2}/2=\hbar^2 \omega_{pi}^2/(m_e c^2)^2$, is the quantum diffraction coefficient which is related to the relativity parameter via $(m_e/m_i)H_r^{2}/2\simeq 4.6\times 10^{-10} R_0^{3}$. On the other hand, the parameter $R_0$ is related also to the mass-density (of white dwarf, for instance) through the relation $\rho\simeq 2m_p n_{e0}(1-\alpha)$ or $\rho(gr/cm^{3})=(1-\alpha)\rho_0 R_0^{3}$ with $\rho_0(gr/cm^{3})\simeq 1.97\times 10^6$, where, $m_p$ is the proton mass. Note that, the density $\rho_0$ is exactly within the range of mass-density of a typical white dwarf and a contribution of $(1-\alpha)$ is included in mass-density definition for the electron-positron pair production/anihilation effect. The density of typical white dwarfs can be in the range $10^5<\rho(gr/cm^{3})<10^{8}$, which, neglecting electron-positron pair production/anihilation, results in values of $0.37<R_0<8$ for the relativity parameter.

\textbf{Before proceeding with calculations, a clear definition of the relativistic degeneracy and distinction between a low-pressure relativistic plasma from the relativistically degenerate quantum Fermi-gas is in order. The relativistic degeneracy is a completely quantum phenomenon ruled by the uncertainty principle and is increased due to the decrease in inter-fermion distances in degenerated plasmas. Although the relativistic effects arise due to increase in fermion number-density in a dense degenerate plasma, however, unlike for the low-pressure relativistic plasmas the degeneracy pressure in the fermion momentum fluid equation usually dominates the relativistic dynamic effects in super-dense plasma state. Chandrasekhar \cite{chandra1}, combining the relativity and the quantum statistics, showed that for dense degenerate Fermi-gas such as a white-dwarf with a mass-density, $\rho$, the degeneracy pressure turns from $P_d\propto \rho^{5/3}$ (with polytropic index $3$) dependence for normal degeneracy for the limit $R_0 \rightarrow 0$ to $P_d\propto \rho^{4/3}$ (with polytropic index $2/3$) dependence for ultra-relativistic degeneracy case in the limit $R_0 \rightarrow \infty$. The relativistic degeneracy starts at mass density of about $4.19\times 10^6 (gr/cm^3)$ of the order in the core a $0.3M_\odot$ white dwarf, which corresponds to a Fermi-momentum $P_{Fe}\sim 1.29 m_e c$ (corresponding to the relativistic degeneracy parameter value of $R_0\sim 1.29$) or the threshold velocity of $u_{Fe}\sim 0.63c$ (the Fermi relativistic factor $\gamma_{Fe}\sim 1.287$). Now comparing the terms in the momentum equation Eq. (\ref{normal}) reveals that the term containing mass ratio is still negligible compared to the degeneracy pressure term. The similar treatment of ultrarelativistic quantum plasma as in our case has been considered in Refs. \cite{rasheed, mamun}. However, this assumption breaks down for a relativistic low-pressure plasma and one has to follow a coherent relativistic formalism given in Refs. \cite{nam, trib}. Thus, for the range of the relativistic degeneracy parameter $R_0^{3}\sim \rho/\rho_0 <10$, which is used in our analysis (see discussion), it is safe to neglect the relativistic dynamic effects compared to that of the dominant relativistic degeneracy pressure. On the other hand, it should be noted that the current analysis includes the whole range of degeneracy parameter, $R_0$, and is not limited only to the relativistic degeneracy range.}

The normalized set of QHD equations can be obtained from the following scaling
\begin{equation}\label{T}
\nabla  \to \frac{{{c_{sr}}}}{{{\omega _{pi}}}}\bar \nabla,\hspace{3mm}t \to \frac{{\bar t}}{{{\omega _{pi}}}},\hspace{3mm}n \to {n_0}\bar n,\hspace{3mm}{\bf{V}} \to {c_{sr}}{\bf{\bar V}},\hspace{3mm}\phi  \to \frac{{{m_e}{c^2}}}{e}\bar \phi .
\end{equation}
where, $c$, ${\omega _{pi}} = \sqrt {4\pi {e^2}{n_{e0}}/{m_i}}$ and ${c_{sr}} = \sqrt {{m_e} {c^2}/{m_i }}$ are the vacuum light speed and the characteristic plasma frequency and relativistic sound-speed, respectively, and the parameter $n_{e0}$ denotes the electron equilibrium density. Now, neglecting the terms containing the small mass-ratio, $m_e/m_i$ and assuming that $k_B T_i \ll m_e c^2$, we arrive at the following simplified set of basic normalized equations
\begin{equation}\label{comp}
\begin{array}{l}
\frac{{\partial {n_i}}}{{\partial t}} + \nabla  \cdot \left( {{n_i}{{\bf{V}}_i}} \right) = 0, \\
\frac{{\partial {n_j}}}{{\partial t}} + \nabla  \cdot \left( {{n_j}{{\bf{V}}_j}} \right) = 0, \\
\frac{{\partial {{\bf{V}}_i}}}{{\partial t}} + \left( {{{\bf{V}}_i} \cdot \nabla } \right){{\bf{V}}_i} = - \nabla \phi + \bar \omega ({{\bf{V}}_i} \times {\bf{k}})=0, \\
{{{s_j}}}\nabla \phi = - \nabla\left[ {1 + {R_0^{2}\alpha_j^{2/3} {n_j}^{2/3}}} \right]^{1/2},\\
{\nabla ^2}\phi = - {\sum\limits_j {{s_j}{n_j}}  - {n_i}}, \\
\end{array}
\end{equation}
where, $\overline{\omega}=\omega_{ci} / \omega_{pi}$ with the ion-cyclotron frequency of $\omega_{ci}=eB_{0} /m_{i}$. Solving Eqs. (\ref{comp}) for the electron and positron number-densities in terms of electrostatic potential with the appropriate boundary requirements ($\mathop {\lim }\limits_{{v_j} \to 0} {n_j} = \alpha_j$ and $\mathop {\lim }\limits_{{v_j} \to 0} \phi  = 0$), results in the following energy relations
\begin{equation}\label{phis}
\begin{array}{l}
{n_e} = R_0^{ - 3}{\left[ {R_0^2 + \phi \left( {2\sqrt {1 + R_0^2}  + \phi } \right)} \right]^{3/2}} \\
{n_p} = {\alpha ^{ - 1}}R_0^{ - 3}{\left[ {R_0^2{\alpha ^{4/3}} - \phi (2\sqrt {1 + R_0^2{\alpha ^{4/3}}}  - \phi )} \right]^{3/2}}, \\
\end{array}
\end{equation}
where, the equilibrium charge neutrality condition is given by Poisson's relation as
\begin{equation}
\alpha  + \beta  = 1,\hspace{3mm}\alpha  = \frac{{{n_{p0}}}}{{{n_{e0}}}},\hspace{3mm}\beta  = \frac{{{n_{i0}}}}{{{n_{e0}}}},
\end{equation}
Transforming the normalized plasma equations (Eqs. (\ref{normal})) to the appropriate strained coordinate defined below admits successful separation of variables and allows elimination of secular terms, which consequently, leads to the desired evolution equations and the corresponding collision parameters \cite{jeffery, washimi, masa}
\begin{equation}\label{stretch}
\begin{array}{l}
\xi  = \varepsilon (kx + ly + mz - {c_\xi }t) + {\varepsilon ^2}{P_0}(\eta ,\tau ) + {\varepsilon ^3}{P_1}(\xi ,\eta ,\tau ) +  \ldots , \\
\eta  = \varepsilon (kx + ly + mz - {c_\eta }t) + {\varepsilon ^2}{Q_0}(\xi ,\tau ) + {\varepsilon ^3}{Q_1}(\xi ,\eta ,\tau ) +  \ldots , \\
\tau  = {\varepsilon ^3}t,\hspace{3mm}{c_\xi } = c,\hspace{3mm}{c_\eta } =  - c, \\
\end{array}
\end{equation}
where, the interrelated functions $P_l$ and $Q_l$ ($l=0,1,2,...$) denote the phase of the traveling solitary waves to be determined together with the evolution equations in next section. The solitons are assumed to travel at directions described by cosine indices $(k,l,m)$ and the direction of magnetic field with respect to the collision line is presented by $\gamma$ angle which is determined in the following relation
\begin{equation}
\begin{array}{l}
m = \cos\gamma,\\
k^2 + l^2 + m^2 = 1.\\
\end{array}
\end{equation}
Now, expanding the dependent plasma variables around equilibrium state through the smallness, $\varepsilon$ parameter which is of the order of perturbation amplitude being a measure of nonlinearity strength \cite{infeld}, we get
\begin{equation}\label{Ordering}
\left[ {\begin{array}{*{20}{c}}
{{n_i}}  \\
{\begin{array}{*{20}{c}}
{{u_i}}  \\
{{v_i}}  \\
{{w_i}}  \\
\end{array}}  \\
\varphi   \\
\end{array}} \right] = \left[ {\begin{array}{*{20}{c}}
1-\alpha   \\
{\begin{array}{*{20}{c}}
0  \\
0  \\
0  \\
\end{array}}  \\
0  \\
\end{array}} \right] + {\varepsilon ^2}\left[ {\begin{array}{*{20}{c}}
{n_i^{(1)}}  \\
{\begin{array}{*{20}{c}}
0  \\
0  \\
{w_i^{(1)}}  \\
\end{array}}  \\
{{\varphi ^{(1)}}}  \\
\end{array}} \right] + {\varepsilon ^3}\left[ {\begin{array}{*{20}{c}}
{n_i^{(2)}}  \\
{\begin{array}{*{20}{c}}
{u_i^{(1)}}  \\
{v_i^{(1)}}  \\
{w_i^{(2)}}  \\
\end{array}}  \\
{{\varphi ^{(2)}}}  \\
\end{array}} \right] + {\varepsilon ^4}\left[ {\begin{array}{*{20}{c}}
{n_i^{(3)}}  \\
{\begin{array}{*{20}{c}}
{u_i^{(2)}}  \\
{v_i^{(2)}}  \\
{w_i^{(3)}}  \\
\end{array}}  \\
{{\varphi ^{(3)}}}  \\
\end{array}} \right] +  \ldots
\end{equation}
The reduced set of plasma equations in new strained coordinate is presented in appendix A. From the lowest-orders in $\varepsilon$ we obtain the following relations
\begin{subequations}\label{leading}
\begin{equation}
c\left( {-\frac{\partial }{{\partial \xi }} + \frac{\partial }{{\partial \eta }}} \right)n_i^{(1)} + m(1-\alpha) \left( {\frac{\partial }{{\partial \xi }} + \frac{\partial }{{\partial \eta }}} \right)w_i^{(1)} = 0,
\end{equation}
\begin{equation}
k\left( {\frac{\partial }{{\partial \xi }} + \frac{\partial }{{\partial \eta }}} \right){\varphi ^{(1)}} - \bar \omega v_i^{(1)} = 0,
\end{equation}
\begin{equation}
l\left( {\frac{\partial }{{\partial \xi }} + \frac{\partial }{{\partial \eta }}} \right){\varphi ^{(1)}} + \bar \omega u_i^{(1)} = 0,
\end{equation}
\begin{equation}
c\left( { - \frac{\partial }{{\partial \xi }} + \frac{\partial }{{\partial \eta }}} \right)w_i^{(1)} + m\left( {\frac{\partial }{{\partial \xi }} + \frac{\partial }{{\partial \eta }}} \right){\varphi ^{(1)}} = 0,
\end{equation}
\begin{equation}
n_i^{(1)} = 3R_0^{ - 2}\left[ {\sqrt {1 + R_0^2}  + {\alpha ^{ - 1/3}}\sqrt {1 + R_0^2{\alpha ^{4/3}}} } \right]{\varphi ^{(1)}},
\end{equation}
\end{subequations}
which leads to the following first-order approximations of plasma variables
\begin{equation}\label{Firstcomp}
\begin{array}{l}
n_i^{(1)} = 3R_0^{ - 2}\left[ {\sqrt {1 + R_0^2}  + {\alpha ^{ - 1/3}}\sqrt {1 + R_0^2{\alpha ^{4/3}}} } \right]\left[ {\varphi ^{(1)}(\xi ,\tau ) + \varphi ^{(1)}(\eta ,\tau )} \right], \\
u_i^{(1)} =  - \frac{{l}}{{\bar \omega}}\left[ {{\partial _\xi}\varphi ^{(1)}(\xi ,\tau ) + {\partial _\eta }\varphi ^{(1)}(\eta ,\tau )} \right],\hspace{3mm} v_i^{(1)} = \frac{{k}}{{\bar \omega}}\left[ {{\partial _\xi }\varphi ^{(1)}(\xi ,\tau ) + {\partial _\xi }\varphi ^{(1)}(\eta ,\tau )} \right], \\ w_i^{(1)} = \frac{{c}}{{m(1-\alpha) }}3R_0^{ - 2}\left[ {\sqrt {1 + R_0^2}  + {\alpha ^{ - 1/3}}\sqrt {1 + R_0^2{\alpha ^{4/3}}} } \right]\left[ {\varphi ^{(1)}(\xi ,\tau ) - \varphi ^{(1)}(\eta ,\tau )} \right],\\
\end{array}
\end{equation}
On the other hand, the compatibility relation can be written as
\begin{equation}\label{dispers}
\frac{{(1-\alpha) {m^2}}}{{{c^2}}} = 3R_0^{ - 2}\left[ {\sqrt {1 + R_0^2}  + {\alpha ^{ - 1/3}}\sqrt {1 + R_0^2{\alpha ^{4/3}}} } \right],
\end{equation}
with the normalized phase-speed, $c$, given as
\begin{equation}\label{speed}
c = \frac{{{R_0}}}{{\sqrt 3 }}\sqrt {\frac{{1 - \alpha }}{{\sqrt {1 + R_0^2}  + {\alpha ^{ - 1/3}}\sqrt {1 + R_0^2{\alpha ^{4/3}}} }}} \cos \gamma.
\end{equation}
From the next higher-order in $\varepsilon$, we get the second-order approximations for variables which are of the following shapes
\textbf{}
\begin{subequations}\label{second}
\begin{equation}
\begin{array}{l}
c\left( { - \frac{\partial }{{\partial \xi }} + \frac{\partial }{{\partial \eta }}} \right)n_i^{(2)} + k\left( {\frac{\partial }{{\partial \xi }} + \frac{\partial }{{\partial \eta }}} \right)u_i^{(1)} +  \\
l\left( {\frac{\partial }{{\partial \xi }} + \frac{\partial }{{\partial \eta }}} \right)v_i^{(1)} + m\left( {\frac{\partial }{{\partial \xi }} + \frac{\partial }{{\partial \eta }}} \right)w_i^{(2)} = 0, \\
\end{array}
\end{equation}
\begin{equation}
\begin{array}{l}
c\left( { - \frac{\partial }{{\partial \xi }} + \frac{\partial }{{\partial \eta }}} \right)u_i^{(1)} + k\left( {\frac{\partial }{{\partial \xi }} + \frac{\partial }{{\partial \eta }}} \right){\varphi ^{(2)}} - \bar \omega v_i^{(2)} = 0, \\
\end{array}
\end{equation}
\begin{equation}
\begin{array}{l}
c\left( { - \frac{\partial }{{\partial \xi }} + \frac{\partial }{{\partial \eta }}} \right)v_i^{(1)} + l\left( {\frac{\partial }{{\partial \xi }} + \frac{\partial }{{\partial \eta }}} \right){\varphi ^{(2)}} + \bar \omega u_i^{(2)} = 0, \\
\end{array}
\end{equation}
\begin{equation}
\begin{array}{l}
c\left( { - \frac{\partial }{{\partial \xi }} + \frac{\partial }{{\partial \eta }}} \right)w_i^{(2)} + m\left( {\frac{\partial }{{\partial \xi }} + \frac{\partial }{{\partial \eta }}} \right){\varphi ^{(2)}} = 0,
\end{array}
\end{equation}
\begin{equation}
n_i^{(2)} = 3R_0^{ - 2}\left[ {\sqrt {1 + R_0^2}  + {\alpha ^{ - 1/3}}\sqrt {1 + R_0^2{\alpha ^{4/3}}} } \right]{\varphi ^{(2)}},
\end{equation}
\end{subequations}
yielding the following second-order approximates
\begin{equation}\label{uv2}
\begin{array}{l}
n_i^{(2)} = 3R_0^{ - 2}\left[ {\sqrt {1 + R_0^2}  + {\alpha ^{ - 1/3}}\sqrt {1 + R_0^2{\alpha ^{4/3}}} } \right]\left[ {\varphi ^{(2)}(\xi ,\tau ) + \varphi ^{(2)}(\eta ,\tau )} \right], \\
u_i^{(2)} = \frac{{ck}}{{{{\bar \omega }^2}}}\left[ {{\partial _\xi }\varphi ^{(2)}(\xi ,\tau ) - {\partial _\eta }\varphi ^{(2)}(\eta ,\tau )} \right] - \frac{{l}}{{\bar \omega}}\left[ {{\partial _{\xi \xi }}\varphi ^{(1)}(\xi ,\tau ) + {\partial _{\eta \eta }}\varphi ^{(1)}(\eta ,\tau )} \right], \\
v_i^{(2)} = \frac{{cl}}{{{{\bar \omega }^2}}}\left[ {{\partial _\xi }\varphi ^{(2)}(\xi ,\tau ) - {\partial _\eta }\varphi ^{(2)}(\eta ,\tau )} \right] + \frac{{k}}{{\bar \omega}}\left[ {{\partial _{\xi \xi }}\varphi ^{(1)}(\xi ,\tau ) + {\partial _{\eta \eta }}\varphi ^{(1)}(\eta ,\tau )} \right], \\
w_i^{(2)} = \frac{{c}}{{m(1-\alpha) }}3R_0^{ - 2}\left[ {\sqrt {1 + R_0^2}  + {\alpha ^{ - 1/3}}\sqrt {1 + R_0^2{\alpha ^{4/3}}} } \right]\left[ {\varphi ^{(2)}(\xi ,\tau ) - \varphi ^{(2)}(\eta ,\tau )} \right]. \\
\end{array}
\end{equation}
where, ${\varphi ^{(1)}}(\xi,\tau)$ and ${\varphi ^{(1)}}(\eta,\tau)$ describe the first-order amplitude evolution and ${\varphi ^{(2)}}(\xi,\tau)$ and ${\varphi ^{(2)}}(\eta,\tau)$ describe the second-order amplitude evolution components of two distinct solitary excitations in the oblique directions ${\eta_ \bot }$ and ${\xi_ \bot }$ (${\eta_ \bot=-\xi_ \bot }$), respectively. In what follows, we will always employ the notations ${\varphi_\xi ^{(1)}}$ and ${\varphi_\eta ^{(1)}}$ instead of ${\varphi ^{(1)}}(\xi,\tau)$ and ${\varphi ^{(1)}}(\eta,\tau)$ for clarity.

\section{Soliton Dynamics and Collision Parameters}\label{shift}

Proceeding to the next higher-order approximation, again by solving the coupled differential equations in this approximation level and by making use of compatibility relation (Eq. \ref{dispers}) and the previous plasma approximations, we obtain
\begin{equation}\label{n3}
\begin{array}{l}
n_i ^{(3)} =\frac{N}{4}\left[ {\frac{{\partial \varphi _\eta^{(1)}}}{{\partial \tau}} + A\varphi _\eta^{(1)}\frac{{\partial \varphi _\eta^{(1)}}}{{\partial \eta}} - B\frac{{{\partial ^3}\varphi _\eta^{(1)}}}{{\partial {\eta^3}}}} \right]\xi -
\frac{N}{4}\left[ {\frac{{\partial \varphi _\xi^{(1)}}}{{\partial \tau}} - A\varphi _\xi^{(1)}\frac{{\partial \varphi _\xi^{(1)}}}{{\partial \xi}} + B\frac{{{\partial ^3}\varphi _\xi^{(1)}}}{{\partial {\xi^3}}}} \right]\eta +  \\
{\frac{E_2}{4}}\left[ {{P_0}(\eta,\tau) + \frac{E_1}{E_2}\int {\varphi _\eta^{(1)}d\eta} } \right]\frac{{\partial \varphi _\xi^{(1)}}}{{\partial \xi}} -{\frac{E_2}{4}}\left[ {{Q_0}(\xi,\tau) - \frac{E_1}{E_2}\int {\varphi _\xi^{(1)}d\xi} } \right]\frac{{\partial \varphi _\eta^{(1)}}}{{\partial \eta}} +  \\ \frac{N}{4}\left[ {\int {\frac{{\partial \varphi _\xi^{(1)}}}{{\partial \tau}}d\xi}  - \int {\frac{{\partial \varphi _\eta^{(1)}}}{{\partial \tau}}d\eta}}\right] - \frac{E_1}{8}\left[ {{{(\varphi _\xi^{(1)})}^2} - {{(\varphi _\eta^{(1)})}^2}} \right] - \frac{E_1}{4}\left[ {\frac{{{\partial ^2}\varphi _\xi^{(1)}}}{{\partial {\xi^2}}} - \frac{{{\partial ^2}\varphi _\eta^{(1)}}}{{\partial {\eta^2}}}} \right] + \\
F(\xi,\tau) + G(\eta,\tau), \\
\end{array}
\end{equation}
where, the entities $F(\xi,\tau)$ and $G(\eta,\tau)$ are assumed to be the homogenous solutions of differential equations. The coefficients in Eq. (\ref{n3}) are as follows
\begin{subequations}\label{coeffs}
\begin{equation}
A = \frac{{\left( {1 + 2R_0^{2}\left( {1 - {\alpha ^{ - 1/3}}} \right) - {\alpha ^{ - 5/3}}} \right)}}{{4\sqrt 3 (1 - \alpha ){R_0}}}{\left[ {\frac{{1 - \alpha }}{{\sqrt {1 + R_0^{2}}  + {\alpha ^{ - 1/3}}\sqrt {1 + R_0^{2}{\alpha ^{4/3}}} }}} \right]^{3/2}}\cos \gamma,
\end{equation}
\begin{equation}
B = \frac{{R_0^3\left( {{\bar \omega ^2} + (1 - \alpha ){{\sin }^2}\gamma } \right)}}{{6\sqrt 3 (1 - \alpha ){\bar \omega ^2}}}{\left[ {\frac{{1 - \alpha }}{{\sqrt {1 + R_0^2}  + {\alpha ^{ - 1/3}}\sqrt {1 + R_0^2{\alpha ^{4/3}}} }}} \right]^{3/2}}\cos \gamma,
\end{equation}
\begin{equation}
N = 6\sqrt 3 (1 - \alpha )R_0^3{\left[ {\frac{{1 - \alpha }}{{\sqrt {1 + R_0^2}  + {\alpha ^{ - 1/3}}\sqrt {1 + R_0^2{\alpha ^{4/3}}} }}} \right]^{ - 3/2}}\sec \gamma,
\end{equation}
\begin{equation}
E_1 = - 3R_0^{ - 4}\left( {1 + 2R_0^2(1 - {\alpha ^{ - 1/3}}) - {\alpha ^{ - 5/3}}} \right),
\end{equation}
\begin{equation}
{E_2} = 12R_0^{ - 2}\left( {\sqrt {1 + R_0^2}  + {\alpha ^{ - 1/3}}\sqrt {1 + R_0^2{\alpha ^{4/3}}} } \right),
\end{equation}
\end{subequations}
As it is realized already, the two first terms in Eqs. (\ref{n3}) are secular and they diverge at $\xi\rightarrow\pm\infty$ and $\eta\rightarrow\pm\infty$, and therefore must vanish. Thus, we obtain two distinct KdV evolution equations one for each traveling solitary structure. Furthermore, the next two terms in Eq. (\ref{n3}) may become secular \cite{nob, masa} at the next higher-order and they must also vanish. The later conditions determine the collision parameters introduced in Eqs. (\ref{stretch}). However, the full determination soliton dynamics is given by the following coupled differential equations
\begin{equation}\label{kdv1}
\frac{{\partial \varphi _\xi^{(1)}}}{{\partial \tau}} + A\varphi _\xi^{(1)}\frac{{\partial \varphi _\xi^{(1)}}}{{\partial \xi}} - B\frac{{{\partial ^3}\varphi _\xi^{(1)}}}{{\partial {\xi^3}}} = 0,
\end{equation}
\begin{equation}\label{P0}
{P_0}(\eta,\tau) = - \frac{{{E_1}}}{{{E_2}}}\int {\varphi _\eta^{(1)}d\eta},
\end{equation}
\begin{equation}\label{kdv2}
\frac{{\partial \varphi _\eta^{(1)}}}{{\partial \tau}} - A\varphi _\eta^{(1)}\frac{{\partial \varphi _\eta^{(1)}}}{{\partial \eta}} + B\frac{{{\partial ^3}\varphi _\eta^{(1)}}}{{\partial {\eta^3}}} = 0,
\end{equation}
\begin{equation}\label{Q0}
{Q_0}(\xi,\tau) = \frac{{{E_1}}}{{{E_2}}}\int {\varphi _\xi^{(1)}d\xi},
\end{equation}
In order to get single-soliton solutions for Eqs. (\ref{kdv1}) and (\ref{kdv2}), with multi-soliton solutions, we require that the perturbed potential components and their derivatives vanish at infinity, i.e.
\begin{equation}\label{boundary}
\begin{array}{l}
\mathop {\lim }\limits_{\zeta \to \pm\infty } \{\varphi _\zeta^{(1)},\frac{\partial \varphi _\zeta^{(1)}}{\partial \zeta },\frac{\partial ^2\varphi _\zeta^{(1)}}{\partial \zeta
^2}\}=0,\hspace{3mm} \zeta={\xi,\eta}.
\end{array}
\end{equation}
Then, we obtain
\begin{equation}\label{phi-x}
\begin{array}{l}
{\varphi _\xi ^{(1)} = \frac{{{\varphi _{\xi 0}}}}{{\cosh^2(\frac{{\xi  - {u_{\xi 0}}\tau }}{{{\Delta _\xi }}})}},}  \\
{{\varphi _{\xi 0}} = \frac{{3{u_{\xi 0}}}}{{{A}}},{\Delta _\xi } = {{(\frac{{4{B}}}{{{u_{\xi 0}}}})}^{\frac{1}{2}}},}  \\
\end{array}
\end{equation}
\begin{equation}\label{phi-y}
\begin{array}{l}
\varphi _\eta^{(1)} = \frac{{{\varphi _{\eta0}}}}{{\cosh^2(\frac{{\eta + {u_{\eta0}}\tau}}{{{\Delta _\eta}}})}},\\
{\varphi _{\eta0}} = \frac{{3{u_{\eta0}}}}{A},\hspace{3mm} {\Delta _\eta} = {(\frac{{4B}}{{{u_{\eta0}}}})^{\frac{1}{2}}}.
\end{array}
\end{equation}
where, $\varphi _{0}$ and $\Delta$ represent the soliton amplitude and width, respectively, and $u_{0}$ is the relativistic Mach-speed.

Close inspection of Eqs. (\ref{coeffs}), reveals that the KdV coefficients $A$ and $B$ in Eqs. (\ref{kdv1}) and (\ref{kdv2}) change sign at $\gamma=\pi/2$, a critical value which determines weather the solitons are compressive or rarefactive. Moreover, it is remarked that the soliton amplitude is independent of the strength of magnetic field, while it strictly depends on applied field direction, $\cos\gamma$. However, the soliton width depends on both magnitude and direction of the external magnetic field. This is in agreement with previous findings \cite{akbari1, Esfand2}.

The collision phase-shifts of solitary excitations are obtained using Eqs. (\ref{P0}) and (\ref{Q0}) together with the KdV solutions (Eqs. (\ref{phi-x}) and (\ref{phi-y})) as
\begin{equation}\label{phase-x}
\begin{array}{l}
{P_0}(\eta,\tau) = \frac{E_1}{E_2} \varphi _{\eta0} \Delta _\eta\tanh(\frac{{\eta - {u_{\eta0}}\tau}}{{{\Delta _\eta}}}),
\end{array}
\end{equation}
\begin{equation}\label{phase-y}
\begin{array}{l}
{Q_0}(\xi,\tau) = \frac{E'_1}{E'_2} \varphi _{\xi0} \Delta _\xi\tanh(\frac{{\xi + {u_{\xi0}}\tau}}{{{\Delta _\xi}}}).
\end{array}
\end{equation}
We can write, up to order $O(\varepsilon^2)$
\begin{equation}\label{trajectory}
\begin{array}{l}
\xi = \varepsilon ({k}x + {l}y + {m}z + {c}t) - \varepsilon^2\frac{{{E_1}}}{{{E_2}}}{\varphi _{\eta0}}{\Delta _\eta}\tanh (\frac{{\eta - {u_{\eta0}}\tau}}{{{\Delta _\eta}}})+O(\varepsilon^3), \\
\eta = \varepsilon ({k}x + {l}y + {m}z - {c}t)] - \varepsilon^2\frac{{{{E'}_1}}}{{{{E'}_2}}}{\varphi _{\xi0}}{\Delta _\xi}\tanh (\frac{{\xi + {u_{\xi0}}\tau}}{{{\Delta _\xi}}})+O(\varepsilon^3), \\
\end{array}
\end{equation}

We can also calculate the overall phase-shifts by comparing the phases of each wave long before and after the collision event in the following form
\begin{equation}
\begin{array}{l}
\Delta {P_0} = P_{post-collision}-P_{past-collision}=\\ \mathop {\lim }\limits_{\xi=0,\eta \to  + \infty } [\varepsilon ({k}x + {l}y + {m}z + {c}t)]-
\mathop {\lim }\limits_{\xi=0,\eta \to  - \infty } [\varepsilon ({k}x + {l}y + {m}z + {c}t)] , \\
\Delta {Q_0} = Q_{post-collision}-Q_{past-collision}=\\ \mathop {\lim }\limits_{\eta=0,\xi \to  + \infty } [\varepsilon ({k}x + {l}y + {m}z - {c}t)]-
\mathop {\lim }\limits_{\eta=0,\xi \to  - \infty } [\varepsilon ({k}x + {l}y + {m}z - {c}t)] , \\
\end{array}
\end{equation}
The phase parameters, $\Delta {P_0}$ and $\Delta {Q_0}$, therefore, present the overall phase-shifts of solitary structures labeled $"S1"$ and $"S2"$ in the head-on collision. Finally, making use of Eqs. (\ref{phase-x}), (\ref{phase-y}) and (\ref{stretch}), we obtain the following simplified expressions
\begin{equation}\label{shifts}
\begin{array}{l}
\Delta {P_0} = {\varepsilon ^2}\left[\frac{{1 + 2R_0^2(1 - {\alpha ^{ - 1/3}}) - {\alpha ^{ - 5/3}}}}{{2R_0^2\left( {\sqrt {1 + R_0^2}  + {\alpha ^{ - 1/3}}\sqrt {1 + R_0^2{\alpha ^{4/3}}} } \right)}}\right]{\varphi _{\eta 0}}{\Delta _\eta }, \\
\Delta {Q_0} = - {\varepsilon ^2}\left[\frac{{1 + 2R_0^2(1 - {\alpha ^{ - 1/3}}) - {\alpha ^{ - 5/3}}}}{{2R_0^2\left( {\sqrt {1 + R_0^2}  + {\alpha ^{ - 1/3}}\sqrt {1 + R_0^2{\alpha ^{4/3}}} } \right)}}\right]{\varphi _{\xi 0}}{\Delta _\xi }. \\
\end{array}
\end{equation}

\section{Numerical Analysis}\label{discussion}

In this section we evaluate the nonlinear wave dynamics using typical values of mass-density for a degenerate white dwarf. Figure 1 depicts the variation of amplitude and width of a solitary excitations with respect to various parameters in a magnetized degenerate plasma with relativistically degenerate electrons and positrons. It is remarked from Figs. 1(a) and 1(b) that, the increase of the mass-density enhances both the amplitude and width of the excitations for fixed other plasma parameters. It is also indicated that the only compressive solitons are present in this plasma for all given parameters. However, regarding the relative positron density variations, one observes that, the soliton amplitude/width increases/decreases as the relative positron density increases. Moreover, as indicated by Figs 1(c), the amplitude of soliton is minimum when the magnetic field is parallel to the propagation direction and increases as the the field direction approaches to the normal, $\gamma=\pi /2$ position. The soliton width, on the other hand, shows (Fig. 1(d)) a maximum value at field-angles defined by
\begin{equation}\label{gamma}
{\gamma _w} = \arccos \frac{1}{\sqrt{3}}{\left[ {1 + \frac{{{{\bar \omega }^2}}}{{1-\alpha}}} \right]^{\frac{1}{2}}},
\end{equation}
where the maximum value of this quantity itself is ${\gamma _{wm}} \approx {54.73^ \circ }$. This is in complete agreement with the cases presented in Ref. \cite{akbari1} treated with a classical model for ultra-relativistic and ordinary electron-positron-ion plasmas. It is from Eq. (\ref{gamma}) remarked that this maximum width field-angle is not affected by the mass-density of plasma. However, the value of the maximum width increases as the plasma gets denser. Figures 1(e) and 1(f) reveal that, while the amplitude increases monotonically with increases in the relative positron number-density, the width has a maximum value which enhances and tend to lower $\alpha$-values as the plasma mass-density increases.

Figure 2 shows variations of the head-on collision phase-shift for interacting solitons. It is observed that, the head-on collision phase-shift is always positive indicating that the post-collision parts of solitons always move ahead of initial trajectories \cite{akbari6}. The value of collision phase-shift increases with increase of the plasma mass-density, as it is concluded from Fig. 2(a). It is also observed from Fig. 2(b) that this parameter is increased as the magnetic field direction moves from the direction parallel to propagation to normal position, however, this effect gets more pronounced when the mass-density of plasma is higher. On the other hand, the variation of collisional phase-shift with respect to the relative positron concentration, presented in Fig. 2(c) is much similar to that of the width of soliton shown in Fig. 1(f). Furthermore, Fig. 2(d) indicates that the increase in the relative magnitude of magnetic field, $\bar \omega$, has also significant effect on the collision of solitons, so that, the higher is the field strength the lower is the shift in the soliton trajectory in the head-on collisions.

\section{Concluding Remarks}\label{conclusion}

\textbf{Using the extended} Poincar\'{e}-Lighthill-Kuo (PLK) reductive perturbation method we studied the oblique propagation and quasi-elastic head-on collisions of solitary structures in a fully degenerate magnetized electron-positron-ion plasma with relativistic ultra-cold electrons and positrons in the framework of quantum hydrodynamics model. It is observed that, plasma parameters such as mass-density, normalized magnetic field strength, its angle with respect to the soliton propagation and the relative positron number-density play important roles on the propagation as well as collision dynamics of solitary structures in a Fermi-Dirac plasma. Current findings can elucidate new aspects of nonlinear dynamics in dense astrophysical objects \textbf{such as white dwarfs and may as well be applicable} to the laboratory produced inertial-confined fusion plasmas.

\appendix

\section{Normalized equations in strained coordinate}
\begin{equation}\label{strain1}
\begin{array}{l}
{\varepsilon ^2}\frac{{\partial {n_i}}}{{\partial \tau }} - c\frac{{\partial {n_i}}}{{\partial \xi }} - {\varepsilon ^2}c\frac{{\partial {Q_0}}}{{\partial \xi }}\frac{{\partial {n_i}}}{{\partial \eta }} + c\frac{{\partial {n_i}}}{{\partial \eta }} + {\varepsilon ^2}c\frac{{\partial {P_0}}}{{\partial \eta }}\frac{{\partial {n_i}}}{{\partial \xi }} + k\frac{{\partial {n_i}{u_i}}}{{\partial \xi }} +\\ {\varepsilon ^2}k\frac{{\partial {Q_0}}}{{\partial \xi }}\frac{{\partial {n_i}{u_i}}}{{\partial \eta }} + k\frac{{\partial {n_i}{u_i}}}{{\partial \eta }} + {\varepsilon ^2}k\frac{{\partial {P_0}}}{{\partial \eta }}\frac{{\partial {n_i}{u_i}}}{{\partial \xi }} + l\frac{{\partial {n_i}{v_i}}}{{\partial \xi }} + {\varepsilon ^2}l\frac{{\partial {Q_0}}}{{\partial \xi }}\frac{{\partial {n_i}{v_i}}}{{\partial \eta }} +\\ l\frac{{\partial {n_i}{v_i}}}{{\partial \eta }} + {\varepsilon ^2}l\frac{{\partial {P_0}}}{{\partial \eta }}\frac{{\partial {n_i}{v_i}}}{{\partial \xi }} + m\frac{{\partial {n_i}{w_i}}}{{\partial \xi }} + {\varepsilon ^2}m\frac{{\partial {Q_0}}}{{\partial \xi }}\frac{{\partial {n_i}{w_i}}}{{\partial \eta }} + m\frac{{\partial {n_i}{w_i}}}{{\partial \eta }} +\\ {\varepsilon ^2}m\frac{{\partial {P_0}}}{{\partial \eta }}\frac{{\partial {n_i}{w_i}}}{{\partial \xi }} + \ldots  = 0, \\
\end{array}
\end{equation}
\begin{equation}\label{strain2}
\begin{array}{l}
{\varepsilon ^2}\frac{{\partial {u_i}}}{{\partial \tau }} - c\frac{{\partial {u_i}}}{{\partial \xi }} - {\varepsilon ^2}c\frac{{\partial {Q_0}}}{{\partial \xi }}\frac{{\partial {u_i}}}{{\partial \eta }} + c\frac{{\partial {u_i}}}{{\partial \eta }} + {\varepsilon ^2}c\frac{{\partial {P_0}}}{{\partial \eta }}\frac{{\partial {u_i}}}{{\partial \xi }} + k{u_i}\frac{{\partial {u_i}}}{{\partial \xi }} +  \\
{\varepsilon ^2}k{u_i}\frac{{\partial {Q_0}}}{{\partial \xi }}\frac{{\partial {u_i}}}{{\partial \eta }} + k{u_i}\frac{{\partial {u_i}}}{{\partial \eta }} + {\varepsilon ^2}k{u_i}\frac{{\partial {P_0}}}{{\partial \eta }}\frac{{\partial {u_i}}}{{\partial \xi }} + l{v_i}\frac{{\partial {u_i}}}{{\partial \xi }} + {\varepsilon ^2}l{v_i}\frac{{\partial {Q_0}}}{{\partial \xi }}\frac{{\partial {u_i}}}{{\partial \eta }} +  \\
l{v_i}\frac{{\partial {u_i}}}{{\partial \eta }} + {\varepsilon ^2}l{v_i}\frac{{\partial {P_0}}}{{\partial \eta }}\frac{{\partial {u_i}}}{{\partial \xi }} + m{w_i}\frac{{\partial {u_i}}}{{\partial \xi }} + {\varepsilon ^2}m{w_i}\frac{{\partial {Q_0}}}{{\partial \xi }}\frac{{\partial {u_i}}}{{\partial \eta }} + m{w_i}\frac{{\partial {u_i}}}{{\partial \eta }} +  \\
{\varepsilon ^2}m{w_i}\frac{{\partial {P_0}}}{{\partial \eta }}\frac{{\partial {u_i}}}{{\partial \xi }} + k\frac{{\partial \varphi }}{{\partial \xi }} + {\varepsilon ^2}k\frac{{\partial {Q_0}}}{{\partial \xi }}\frac{{\partial \varphi }}{{\partial \eta }} + k\frac{{\partial \varphi }}{{\partial \eta }} + {\varepsilon ^2}k\frac{{\partial {P_0}}}{{\partial \eta }}\frac{{\partial \varphi }}{{\partial \xi }} - \frac{{\bar \omega {v_i}}}{\varepsilon } + \ldots  = 0, \\
\end{array}
\end{equation}
\begin{equation}\label{strain3}
\begin{array}{l}
{\varepsilon ^2}\frac{{\partial {v_i}}}{{\partial \tau }} - c\frac{{\partial {v_i}}}{{\partial \xi }} - {\varepsilon ^2}c\frac{{\partial {Q_0}}}{{\partial \xi }}\frac{{\partial {v_i}}}{{\partial \eta }} + c\frac{{\partial {v_i}}}{{\partial \eta }} + {\varepsilon ^2}c\frac{{\partial {P_0}}}{{\partial \eta }}\frac{{\partial {v_i}}}{{\partial \xi }} + k{u_i}\frac{{\partial {v_i}}}{{\partial \xi }} +  \\
{\varepsilon ^2}k{u_i}\frac{{\partial {Q_0}}}{{\partial \xi }}\frac{{\partial {v_i}}}{{\partial \eta }} + k{u_i}\frac{{\partial {v_i}}}{{\partial \eta }} + {\varepsilon ^2}k{u_i}\frac{{\partial {P_0}}}{{\partial \eta }}\frac{{\partial {v_i}}}{{\partial \xi }} + l{v_i}\frac{{\partial {v_i}}}{{\partial \xi }} + {\varepsilon ^2}l{v_i}\frac{{\partial {Q_0}}}{{\partial \xi }}\frac{{\partial {v_i}}}{{\partial \eta }} +  \\
l{v_i}\frac{{\partial {v_i}}}{{\partial \eta }} + {\varepsilon ^2}l{v_i}\frac{{\partial {P_0}}}{{\partial \eta }}\frac{{\partial {v_i}}}{{\partial \xi }} + m{w_i}\frac{{\partial {v_i}}}{{\partial \xi }} + {\varepsilon ^2}m{w_i}\frac{{\partial {Q_0}}}{{\partial \xi }}\frac{{\partial {v_i}}}{{\partial \eta }} + m{w_i}\frac{{\partial {v_i}}}{{\partial \eta }} +  \\
{\varepsilon ^2}m{w_i}\frac{{\partial {P_0}}}{{\partial \eta }}\frac{{\partial {v_i}}}{{\partial \xi }} + l\frac{{\partial \varphi }}{{\partial \xi }} + {\varepsilon ^2}l\frac{{\partial {Q_0}}}{{\partial \xi }}\frac{{\partial \varphi }}{{\partial \eta }} + l\frac{{\partial \varphi }}{{\partial \eta }} + {\varepsilon ^2}l\frac{{\partial {P_0}}}{{\partial \eta }}\frac{{\partial \varphi }}{{\partial \xi }} + \frac{{\bar \omega {u_i}}}{\varepsilon } + \ldots  = 0, \\
\end{array}
\end{equation}
\begin{equation}\label{strain4}
\begin{array}{l}
{\varepsilon ^2}\frac{{\partial {w_i}}}{{\partial \tau }} - c\frac{{\partial {w_i}}}{{\partial \xi }} - {\varepsilon ^2}c\frac{{\partial {Q_0}}}{{\partial \xi }}\frac{{\partial {w_i}}}{{\partial \eta }} + c\frac{{\partial {w_i}}}{{\partial \eta }} + {\varepsilon ^2}c\frac{{\partial {P_0}}}{{\partial \eta }}\frac{{\partial {w_i}}}{{\partial \xi }} + k{u_i}\frac{{\partial {w_i}}}{{\partial \xi }} +  \\
{\varepsilon ^2}k{u_i}\frac{{\partial {Q_0}}}{{\partial \xi }}\frac{{\partial {w_i}}}{{\partial \eta }} + k{u_i}\frac{{\partial {w_i}}}{{\partial \eta }} + {\varepsilon ^2}k{u_i}\frac{{\partial {P_0}}}{{\partial \eta }}\frac{{\partial {w_i}}}{{\partial \xi }} + l{v_i}\frac{{\partial {w_i}}}{{\partial \xi }} + {\varepsilon ^2}l{v_i}\frac{{\partial {Q_0}}}{{\partial \xi }}\frac{{\partial {w_i}}}{{\partial \eta }} +  \\
l{v_i}\frac{{\partial {w_i}}}{{\partial \eta }} + {\varepsilon ^2}l{v_i}\frac{{\partial {P_0}}}{{\partial \eta }}\frac{{\partial {w_i}}}{{\partial \xi }} + m{w_i}\frac{{\partial {w_i}}}{{\partial \xi }} + {\varepsilon ^2}m{w_i}\frac{{\partial {Q_0}}}{{\partial \xi }}\frac{{\partial {w_i}}}{{\partial \eta }} + m{w_i}\frac{{\partial {w_i}}}{{\partial \eta }} +  \\
{\varepsilon ^2}m{w_i}\frac{{\partial {P_0}}}{{\partial \eta }}\frac{{\partial {w_i}}}{{\partial \xi }} + m\frac{{\partial \varphi }}{{\partial \xi }} + {\varepsilon ^2}m\frac{{\partial {Q_0}}}{{\partial \xi }}\frac{{\partial \varphi }}{{\partial \eta }} + m\frac{{\partial \varphi }}{{\partial \eta }} + {\varepsilon ^2}m\frac{{\partial {P_0}}}{{\partial \eta }}\frac{{\partial \varphi }}{{\partial \xi }} + \ldots  = 0, \\
\end{array}
\end{equation}
\begin{equation}\label{strain5}
\begin{array}{l}
{\varepsilon ^2}\left[ {\frac{{{\partial ^2}\varphi }}{{\partial {\xi ^2}}} + \frac{{{\partial ^2}\varphi }}{{\partial {\eta ^2}}}} \right] - \left\{ {1 - \alpha  - {n_i} + 3R_0^{ - 2}\left[ {\sqrt {1 + R_0^2}  + {\alpha ^{ - 1/3}}\sqrt {1 + R_0^2{\alpha ^{4/3}}} } \right] + } \right. \\
\left. {\frac{{3R_0^{ - 4}}}{2}\left[ {(1 + 2R_0^2) - {\alpha ^{ - 5/3}}(1 + 2R_0^2{\alpha ^{4/3}})} \right]} \right\} +  \ldots  = 0. \\
\end{array}
\end{equation}

\newpage

\textbf{FIGURE CAPTIONS}

\bigskip

Figure 1

\bigskip

(Color online) Variations of soliton amplitude (left column of figure) and width (right column of figure) with respect to magnetic field direction $\gamma$, relative positron concentrations, $\alpha$, and normalized mass-density of plasma. The dash sizes in all plots are appropriately related to the values of the varied parameter. The values of $u_{\xi,0}=u_{\eta,0}=0.1$ and $\varepsilon=0.1$ are used for all plots.

\bigskip

Figure 2

\bigskip

(Color online) Variations of the head-on collision phase-shift with respect to with respect to magnetic field direction $\gamma$, relative positron concentrations, $\alpha$, and normalized mass-density of plasma, when other parameters are fixed. The values of $u_{\xi,0}=u_{\eta,0}=0.1$ and $\varepsilon=0.1$ are used for all plots. The dash sizes in all plots are appropriately related to the values of the varied parameter.

\newpage

\begin{figure}[ptb]\label{Figure1}
\includegraphics[scale=.6]{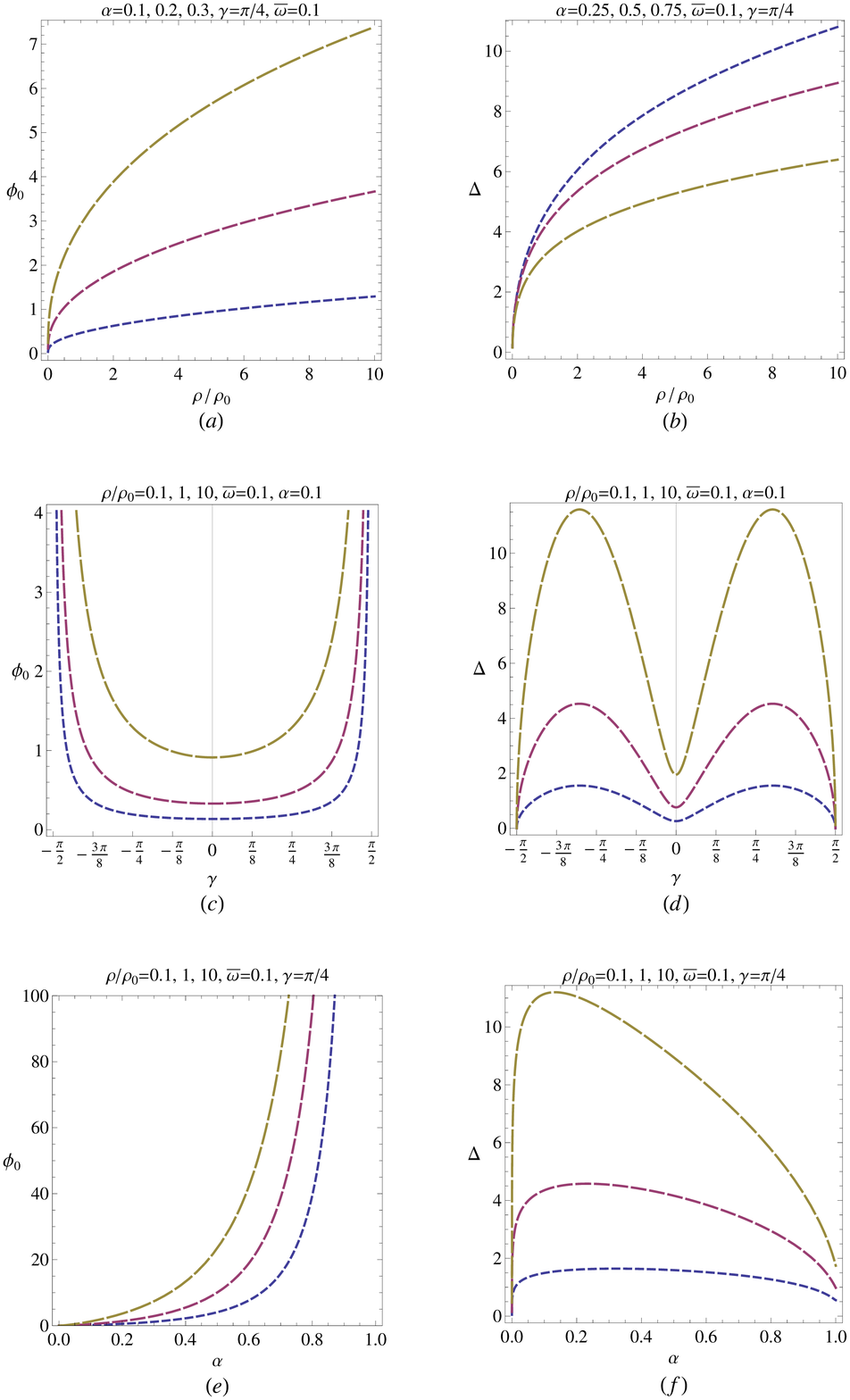}\caption{}
\end{figure}

\newpage

\begin{figure}[ptb]\label{Figure2}
\includegraphics[scale=.6]{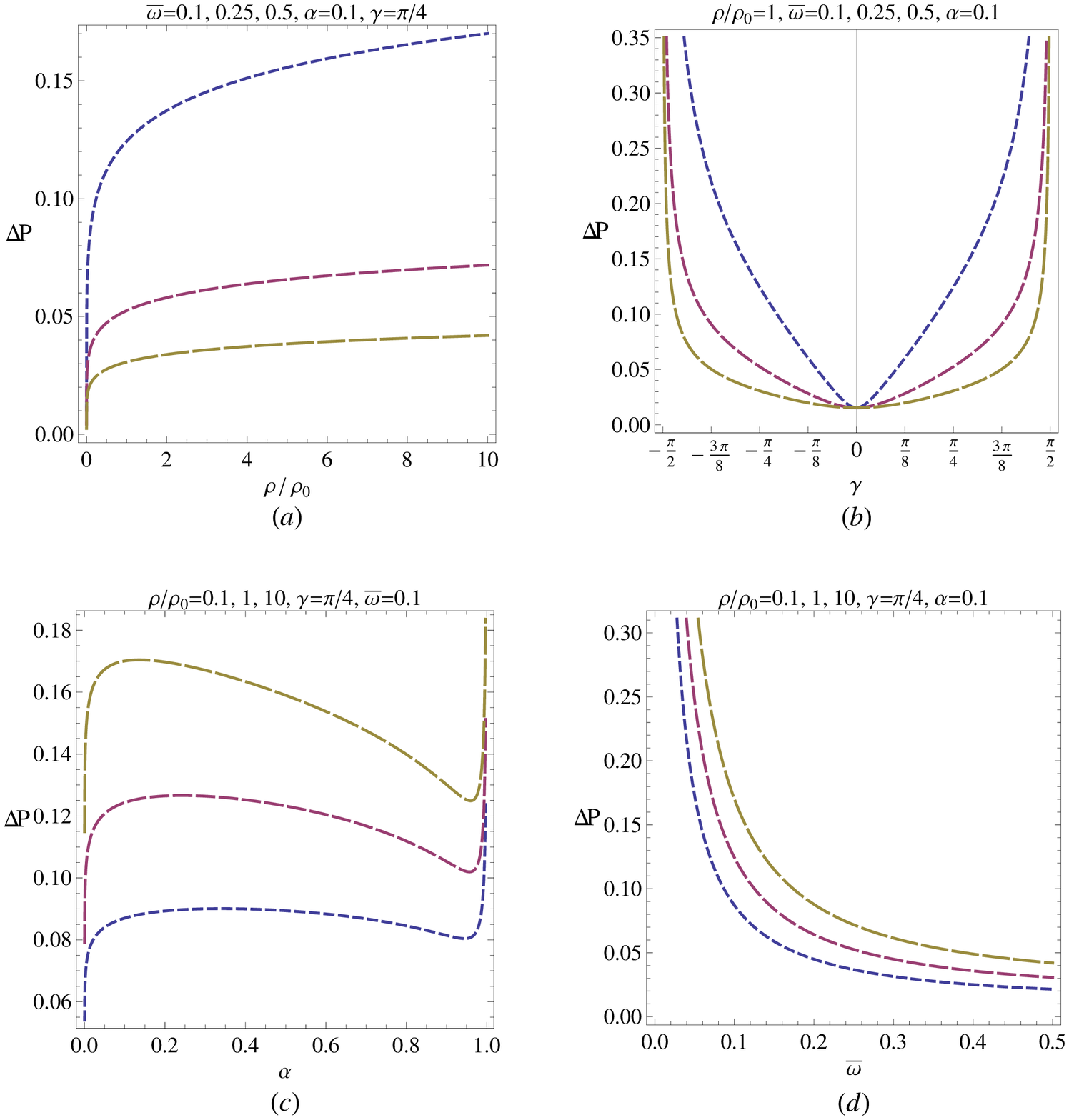}\caption{}
\end{figure}

\newpage

\end{document}